\documentclass[12pt]{iopart}

\usepackage{graphicx}
\begin{document}

\title[Investigation of a $^{85}$Rb DMOT using an ONF]{Investigation of a $^{85}$Rb Dark Magneto-Optical Trap using an Optical Nanofibre}

\author{Laura Russell$^{1,2}$, Ravi Kumar$^{1,2}$, Vibhuti Bhushan Tiwari$^{1,3}$ and S{\`i}le Nic Chormaic$^{1,2,4}$}

\address{$^1$ Department of Physics, University College Cork, Cork, Ireland}
\address{$^2$ Light-Matter Interactions Unit, OIST Graduate University, 1919-1 Tancha, Onna-son, Okinawa 904-0495, Japan}
\address{$^3$ Laser Physics Applications Section, Raja Ramanna Centre for Advanced Technology, Indore 452013, India}
\address{$^4$ School of Chemistry and Physics, University of KwaZulu-Natal, Durban 4001, South Africa}

\ead{ravi.kumar@oist.jp}

\begin{abstract}
We report here measurements on a dark magneto-optical trap (DMOT) of $^{85}$Rb atoms using an optical nanofibre (ONF) with a waist of $\sim$~1 $\mu$m. The DMOT is created using a doughnut-shaped repump beam along with a depump beam for efficient transfer of cold atoms from the bright hyperfine ground state ($F=3$) into the dark hyperfine ground state ($F=2$). The fluorescence from the cold $^{85}$Rb atoms of the DMOT is detected by coupling it into the fibre-guided modes of the ONF. The measured fractional population of cold atoms in the bright hyperfine ground state ($p$) is as low as $\sim$0.04. The dependence of loading rate of DMOT on cooling laser intensity is investigated and also compared with the loading rate of a bright-MOT (BMOT). This work lays the foundation for the use of an ONF for probing of a small number of atoms in an optically-dense cold  atomic cloud.
\end{abstract}

\maketitle

\section{Introduction}
The optical nanofibre (ONF), or optical microfibre depending on the waist diameter, is an ultrathin light guide usually fabricated from commercially-available fibre using a heat-and-pull technique \cite{ward06}. The enhancement of the spontaneous emission from atoms surrounding an ONF into its guided modes is well known \cite{lekien05}, making it an ideal high-sensitivity tool for channeling atomic fluorescence to a detector. In recent years, ONFs have been used as probes for cold atoms \cite{nayak07,morrissey09,russell12,russell13} and hot atomic vapors \cite{spillane08,hendrickson10,watkins13}. A number of proposals for trapping neutral atoms around ONFs have been made \cite{balykin04, fam04, sague08, reitz12, phelan13} and two-colour trapping of caesium has been achieved \cite{vetsch10, goban12}.  Some of the advantages of ONFs are that, in addition to trapping applications, the nanofibre provides an interface that can be exploited for quantum communication using ensembles of laser-cooled atoms or for studying atom-surface interactions \cite{russell09, minogin09}. A review on recent advances in this field is contained in Morrissey \emph{et al.} \cite{morrissey13}.

In the work reported here, atom cloud density enhancement techniques for a sample of laser-cooled $^{85}$Rb atoms are studied in order to increase the fluorescence and/or absorption signals obtainable with ONFs. The density of atoms in a standard magneto-optical trap (MOT) is limited to $\sim10^{11}$ atoms/cm$^3$ for two reasons. First, a transfer of kinetic energy occurs when a ground state atom collides with an excited state atom leading to a trap loss rate that is proportional to the number of atoms trapped. The second limitation is due to laser-cooled atoms in the MOT reabsorbing scattered photons, a phenomenon known as radiation pressure. The outward radiation pressure of the fluorescent light balances the confining forces of the trapping laser beams at a particular density and, at this point, if the number of atoms is increased, the atom cloud simply grows in size rather than in density. For the ONF, this is a major consideration; if one wishes to do absorption experiments using optical nanofibres, as many atoms as possible are required to fill the evanescent region around the fibre in order to optimize  the signal quality.

In 1993 Ketterle \emph{et al.} \cite{ketterle93} demonstrated a MOT configuration, known as the dark MOT (DMOT), which confined sodium atoms in the lower (or dark) hyperfine ground level, where they are unperturbed by the cooling beams and, thus, the DMOT is free from the density limitations of a standard bright MOT (BMOT). By reducing the intensity of the repump beams used in the standard BMOT configuration almost all the atoms are transferred from the bright hyperfine  level, $F=2$, to the dark hyperfine  level, $F=1$, of the $3S_{\frac{1}{2}}$ ground state of Na. Although a small excitation rate is optimal for confining large numbers of atoms at a high density, the maximum possible excitation rate is required to efficiently capture atoms from a vapor and load them into a MOT. By using bright and dark regions in the repump beam simultaneously, both of these requirements can be fulfilled; a doughnut-shaped (or hollow) repump beam can be used for this as it consists of a bright outer ring of light and a dark inner circle. An atom which encounters the dark region (i.e. the region with negligible repump intensity) will only spend a very short time in the cooling cycle before being shelved into the dark hyperfine level. Alternative methods of creating a DMOT have been demonstrated. For example, Muniz \emph{et al.} \cite{muniz04} tuned the frequency of the repump laser to create a DMOT from an intense flux of slowed Na atoms. In the last decade, a more complete understanding of the DMOT has been formed \cite{chapovsky06,chapovsky07,singh10} and recently extended to studies with ultracold atoms and molecules \cite{li-rong12}.

One way of working with a BMOT and DMOT is by generating them sequentially in time using triggered, mechanical shutters and AOMs. However, an alternative approach is to simultaneously generate the BMOT and DMOT in different spatial regions of the trap, as achieved using the aforementioned doughnut-shaped repump beam. The spatially-generated DMOT has  advantages since atoms can be continuously loaded into the dark region from the outer bright MOT where the usual processes occur. In fact, although the inner, dark region is not really a magneto-optical trap, the outer bright region still captures background atoms and feeds the dark section. In other words, experimental parameters for the atoms at the trap center can be monitored and adjusted, while simultaneously maintaining the high loading rate achievable with the bright MOT. This is possible because the trapped atom cloud is localized near the minimum of the magnetic field, whereas the capture of atoms occurs throughout the whole intersection volume of the cooling laser beams. Note that only the spatial DMOT is considered in the work that follows.  This arrangement is very appealing for optical nanofibre-based work. For example, the possibility of continuous absorption measurements via an ONF probe becomes available. Permyakova \emph{et al.} \cite{permyakova08} report that, even with moderate beam powers of the order of 6 mW, the dark MOT can capture a large number of atoms and achieve optical densities as high as 9. Thus, by using low cooling laser intensities (for confining large numbers of atoms at a high density) in a DMOT scheme, high signal-to-noise ratios for the fluorescence coupling should be achievable for minimal beam powers. This leads to the potential demonstration of non-linear optical effects in the atom cloud without requiring high laser powers.

\section{Experimental Technique}

\subsection{Laser configuration}

\begin{figure}[!ht]
\centering
\includegraphics[width=10cm]{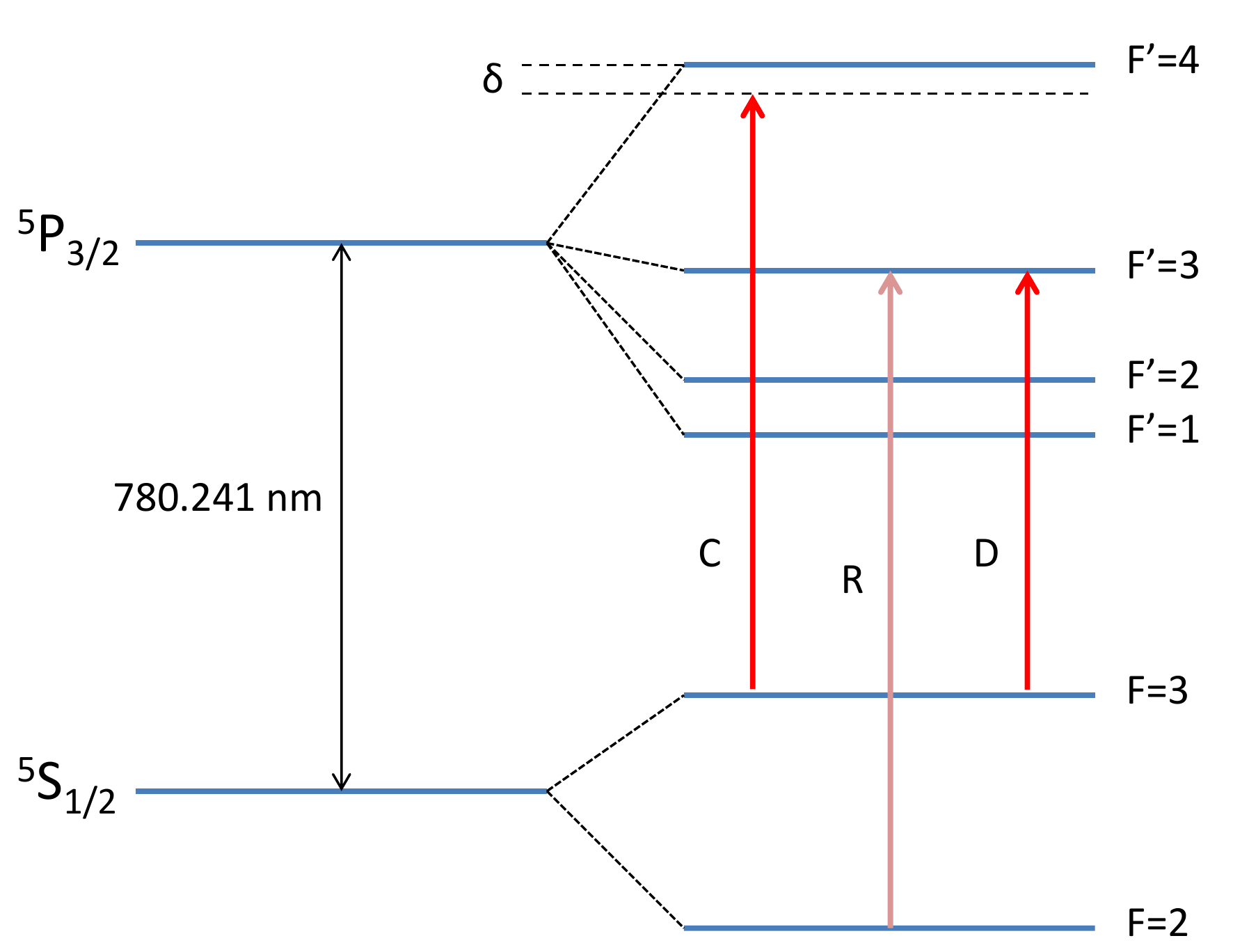}
\caption[Energy levels used in the dark MOT for $^{85}$Rb.]{\label{fig:figure1}Energy levels used in the dark MOT for $^{85}$Rb. `C' represents the cooling laser, `R' represents the repump beam, and `D' represents the depump beam.}
\end{figure}

Earlier, Townsend \emph{et al.} \cite{townsend96} demonstrated atom densities of nearly $10^{12}$ cm$^{-3}$ in a dark MOT for caesium. Prior to this, no report of similar density enhancements in MOTs for elements other than sodium had been reported. Excited state hyperfine splitting is large for heavier alkali-metal atoms which makes it difficult to get the atoms in the dark state, while still maintaining enough atoms in the trap, by transient reduction of repump beam intensity as done in case of Na atoms. Hence, Townsend \emph{et al.} used a technique of partially blocking the repump beam to create a low repump intensity region at the centre surrounded by a higher repump intensity, and directed a depump beam to the dark region to increase the rate of decay to the dark ground state. We follow the same scheme for creating a DMOT for $^{85}$Rb. The depump beam is tuned to the $F=3\rightarrow F'=3$ transition of  $^{85}$Rb. Rb atoms (see \fref{fig:figure1}) undergo many cooling cycles ($F=3\rightarrow F'=4$) before falling to the dark ground state ($F=2$) due to spontaneous Raman processes if no depump beam is used. The use of depump beam accelerates ground state pumping of atoms. 

\begin{figure}[!ht]
\centering
\includegraphics[width=14cm]{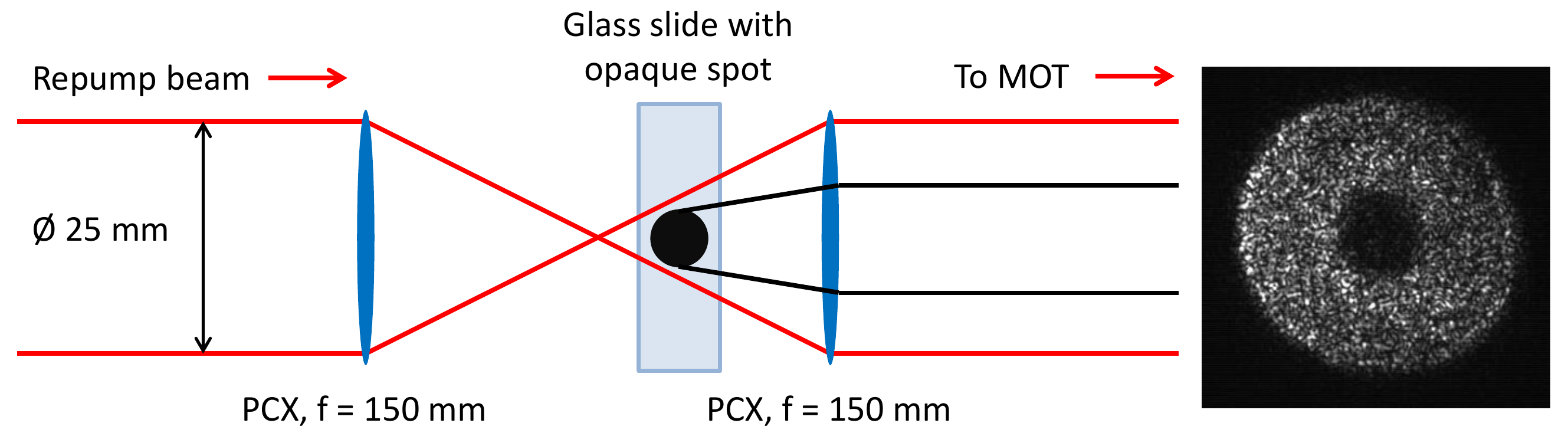}
\caption[Generation of a hollow repump beam]{\label{fig:figure2}Left: Experimental setup used to create the doughnut repump beam using a unity magnifier with an opaque circle positioned near the centre of two lenses. This imaging setup reduces Fresnel diffraction in the beam at the position of the MOT. Right: Repump beam profile obtained using a CCD camera.}
\end{figure}

In the work reported here, a retro-reflected three beam MOT configuration is used for laser-cooling and trapping $^{85}$Rb atoms. The doughnut-shaped repump beam is generated in the following manner. First, the laser beam is expanded in a telescope setup to a diameter of 25 mm (see \fref{fig:figure2}). A separate, unit magnification telescope is used to place a shadow in the beam and project it on to the atom cloud. A glass slide with an opaque circle as the shadow is placed near the focal point of the telescope and aligned to create the doughnut-shaped repump beam at the MOT. The opaque circle has a transmission of $10^{-4}$ as determined experimentally. If the circle were placed in the beam without being incorporated into a telescope system, Fresnel diffraction effects would occur at the trap centre. This would diminish the efficiency of the dark MOT in shelving atoms into  $F=2$. To facilitate  switching from a DMOT to a BMOT, the slide with the shadow in the optical path of the repump beam can be easily removed from its holder, turning the doughnut beam into a standard Gaussian repump beam.

Rather than directing the doughnut repump beam along a path which copropagates with a cooling beam, a separate path is used. The reason for this is due to the retro-reflecting mirrors in the cooling beam path for the MOT design. If the  repump beam were aligned such that it passed centrally through the cloud, but reflected back off a retro-reflecting mirror,  the back-reflected dark spot may not align perfectly with the forward propagating one and could destroy the DMOT.

\subsection{Quantifying the DMOT quality}

To quantify the `darkness' of the trap, Ketterle \emph{et al.} \cite{ketterle93} introduced a quality parameter, $p$, that denotes the fraction of atoms in the bright, ground hyperfine level compared to the total number of atoms in the ground state, where

\begin{equation}
p=\frac{N_b}{N_b+N_d}.
\label{eqn:equation1}
\end{equation}

For $^{85}$Rb, $N_b$ is the number of atoms in the bright $F=3$ level and $N_d$ is the number of atoms in the dark $F=2$ level of the ground state. The quality of the DMOT is determined from $p$, where $p=1$ represents a 100\% bright MOT and $p=0$ is a 100\% dark MOT. For the measurements of $p$ reported here an ONF with diameter $\sim$1 $\mu$m is passed through the centre of the atom cloud. Fluorescence emitted from the laser-cooled atoms couples into the guided modes of the ONF and is subsequently measured at one fibre pigtail end using a single photon counting module (SPCM). The recorded count rate is proportional to the number of atoms in whichever hyperfine ground level is probed, thereby enabling a measurement of $p$ to be performed. Alternatively, $p$ can be estimated from the magnitude of observed absorption dips if free-space spectroscopy is performed.

\section{Results and discussions}

\subsection{Free-space spectroscopy of the DMOT}

\begin{figure}[!ht]
\centering
\includegraphics[width=15cm]{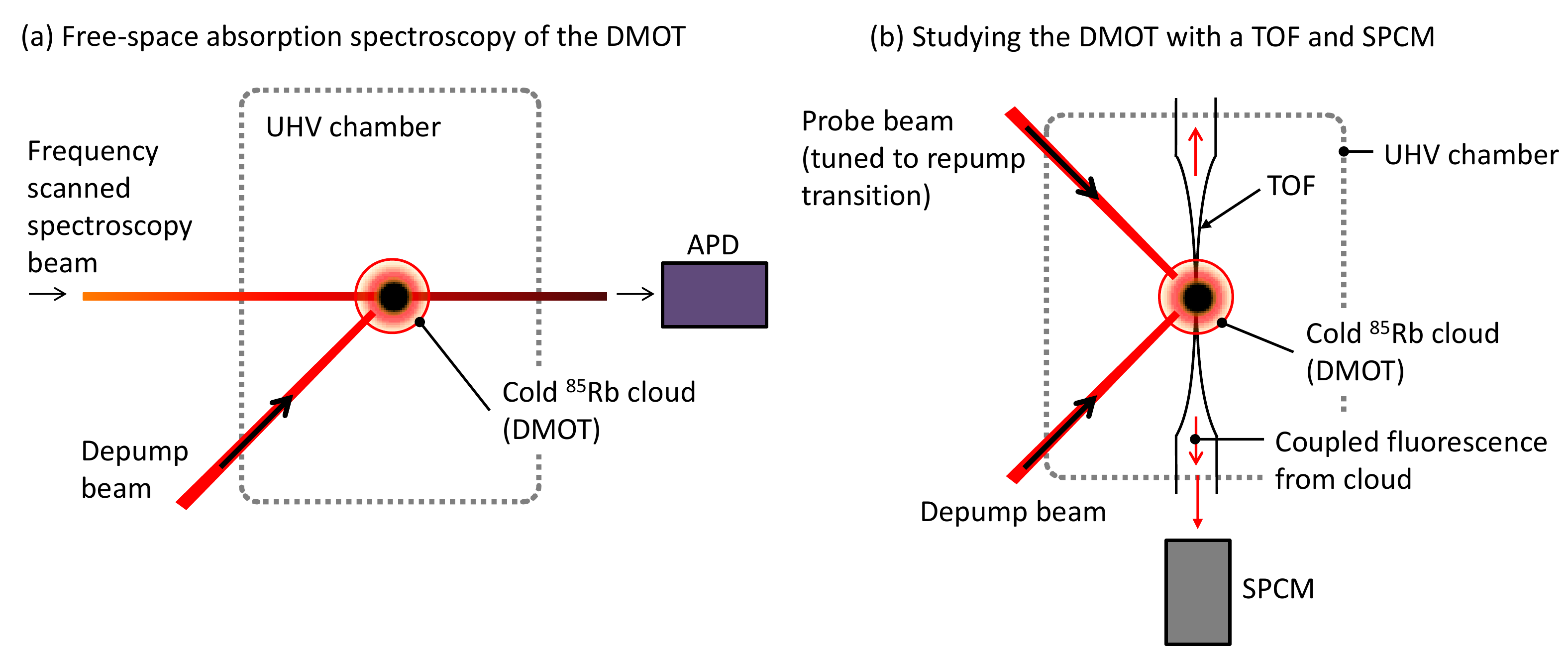}
\caption[Experimental setup for probing a DMOT]{\label{fig:figure3} Schematic of the experimental setup: (a) Free-space spectroscopy on the DMOT incorporating a frequency-scanned probe beam, APD: avalanche photodiode, UHV: ultrahigh vacuum.  (b) Probing the DMOT with an optical fibre, ONF: optical nanofibre, SPCM: single photon counting module.}
\end{figure}

The experimental setup is shown in \fref{fig:figure3}. Initially, a free-space probe beam is passed through the atom cloud for absorption spectroscopy measurements, as illustrated in \fref{fig:figure3}(a). The free-space probe beam is aligned with the centre of the cloud so that it passes through the DMOT. The probe beam is obtained from a 780 nm extended cavity diode laser (ECDL) operating in scanning mode and frequency tuned via saturated absorption spectroscopy. The beam, after passing through the cloud, is directed on an avalanche photodiode (APD) (Hamamatsu model: C5460-01), connected to a computer.

\begin{figure}[!ht]
\centering
\includegraphics[width=15cm]{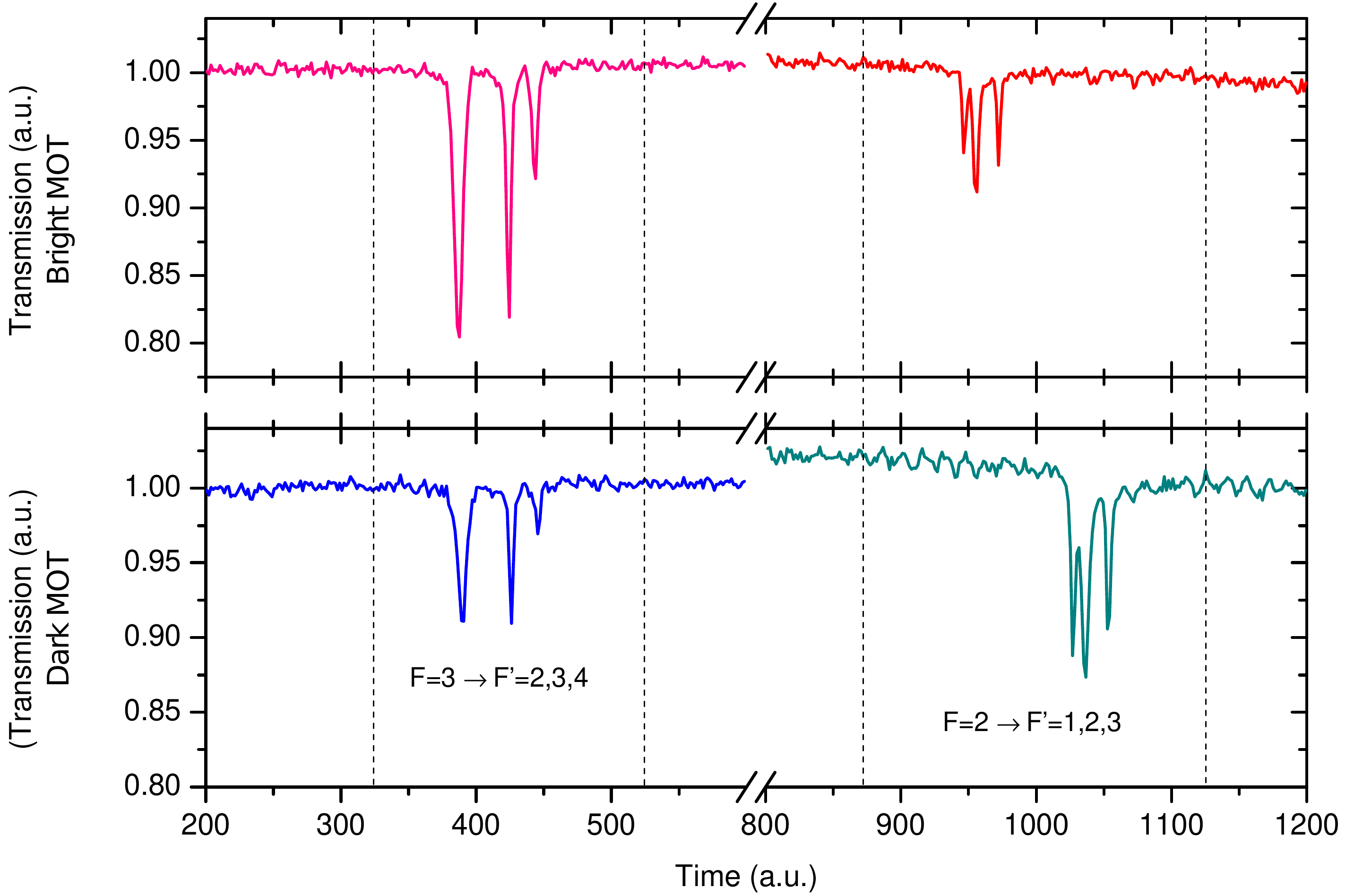}
\caption[Free-space absorption spectroscopy on a BMOT and DMOT]{\label{fig:figure4}Upper trace: Free-space absorption spectroscopy on a BMOT. Lower trace:  Free-space absorption spectroscopy on an optimized DMOT. The DMOT here is created using a doughnut repump MOT beam and a depump beam. There is a higher population in the $F=2$ lower hyperfine level for the DMOT compared to the BMOT.}
\end{figure}

\Fref{fig:figure4} (upper plot) shows the absorption spectrum obtained using the free-space method to obtain a bright MOT. The lower plot in \fref{fig:figure4} shows the spectrum for the DMOT (with the doughnut repump beam and the depump beam). It is evident that the magnitude of the $F=2\rightarrow F'=1,2,3$ dips has doubled for the DMOT compared to the BMOT, while the magnitude of the $F=3\rightarrow F'=2,3,4$ dips has approximately halved, indicating a clear enhancement of the atom population in the $F=2$ hyperfine level for the DMOT compared to the BMOT.

\subsection{Characterizing the DMOT with an ONF}

\begin{figure}[!ht]
\centering
\includegraphics[width=12cm]{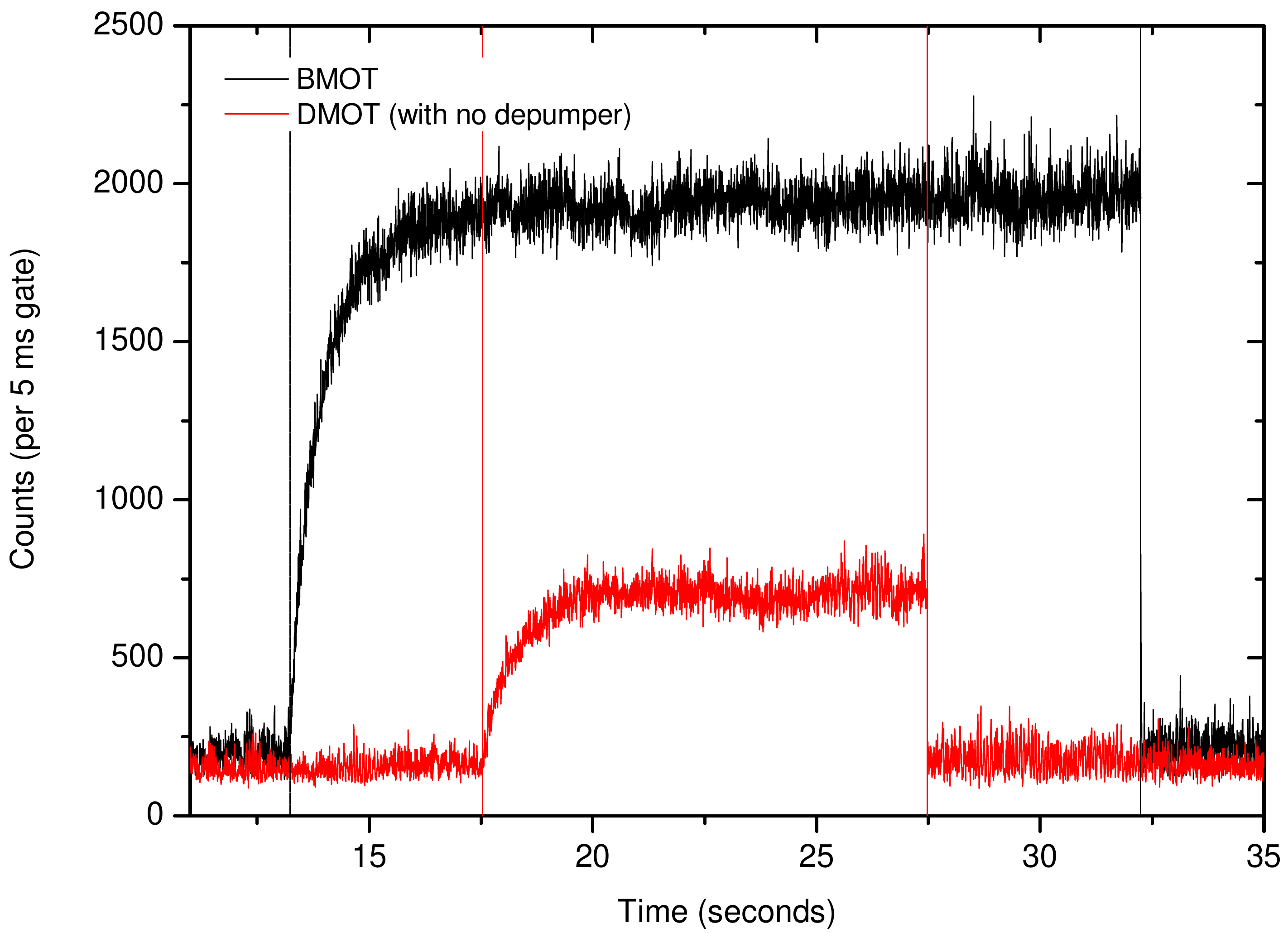}
\caption[Loading curves for a BMOT and DMOT]{\label{fig:figure6}Loading curves for a BMOT (red, upper) and a DMOT (black, lower). This plot shows the residual fluorescence coupling that can be obtained using a DMOT of modest $p$ value. In this case, the DMOT has been created with the doughnut repump beam only (no depump light has been used).}
\end{figure}

By using beam stops to block and unblock the depump beam and/or the probe beam, different populations of atoms can be studied using an ONF. To fabricate the ONF, a heat-and-pull-technique is used \cite{ward06, kenny91, brambilla04}. Commercially-available SM780 (Fibercore) optical fibre is heated using an oxygen-butane flame and pulled using a pair of motorised translational stages. For these experiments, the final transmission of the ONF was $\sim$60\% from end to end and the waist diameter was $\sim$ 1 $\mu$m. After fabrication, the ONF is fixed in a vertical orientation to an aluminium U-shaped mount with UV-curable glue. The nanofibre is aligned centrally in the MOT and the fibre pigtails are coupled into and out of the UHV chamber using a Teflon feedthrough \cite{abraham98,cowpe08}. A single photon counting module (SPCM) is connected to one pigtail of the ONF so that fluorescence from the atom cloud can be recorded. The setup is shown in \fref{fig:figure3}(b). In the previous section, a probe beam generated by an ECDL was used (\fref{fig:figure3}(a)). Now, the probe beam is derived from the repump beam of the MOT and aligned, in free-space, with the center of the atom cloud (\fref{fig:figure3}(b)). The frequency of this probe beam is fixed with an AOM and tuned to the repump transition of $^{85}$Rb. \Fref{fig:figure6} shows two loading curves: the upper (red) plot is for a BMOT and the lower (black) plot is for a DMOT. The DMOT in this instance has no depump shining on it and, thus, it is not extremely dark (i.e. $p$ is not very low). This explains why the resulting coupling from the DMOT ($\sim$750 counts/5 ms) is  not negligible when compared to the coupling from the BMOT ($\sim$2000 counts/5 ms). The steady state count rate for both curves is proportional to the number of atoms in $F=3$ although the distribution of atoms between both hyperfine ground states, $F=2$ and $F=3$, is different in each case. However, these two sets of data are not sufficient to determine the population of each ground state or the darkness of the DMOT. To determine the value of $p$ with the SPCM and ONF two sets of data must be recorded and analyzed, as described in the following section.

\begin{figure}[!ht]
\centering
\includegraphics[width=15cm]{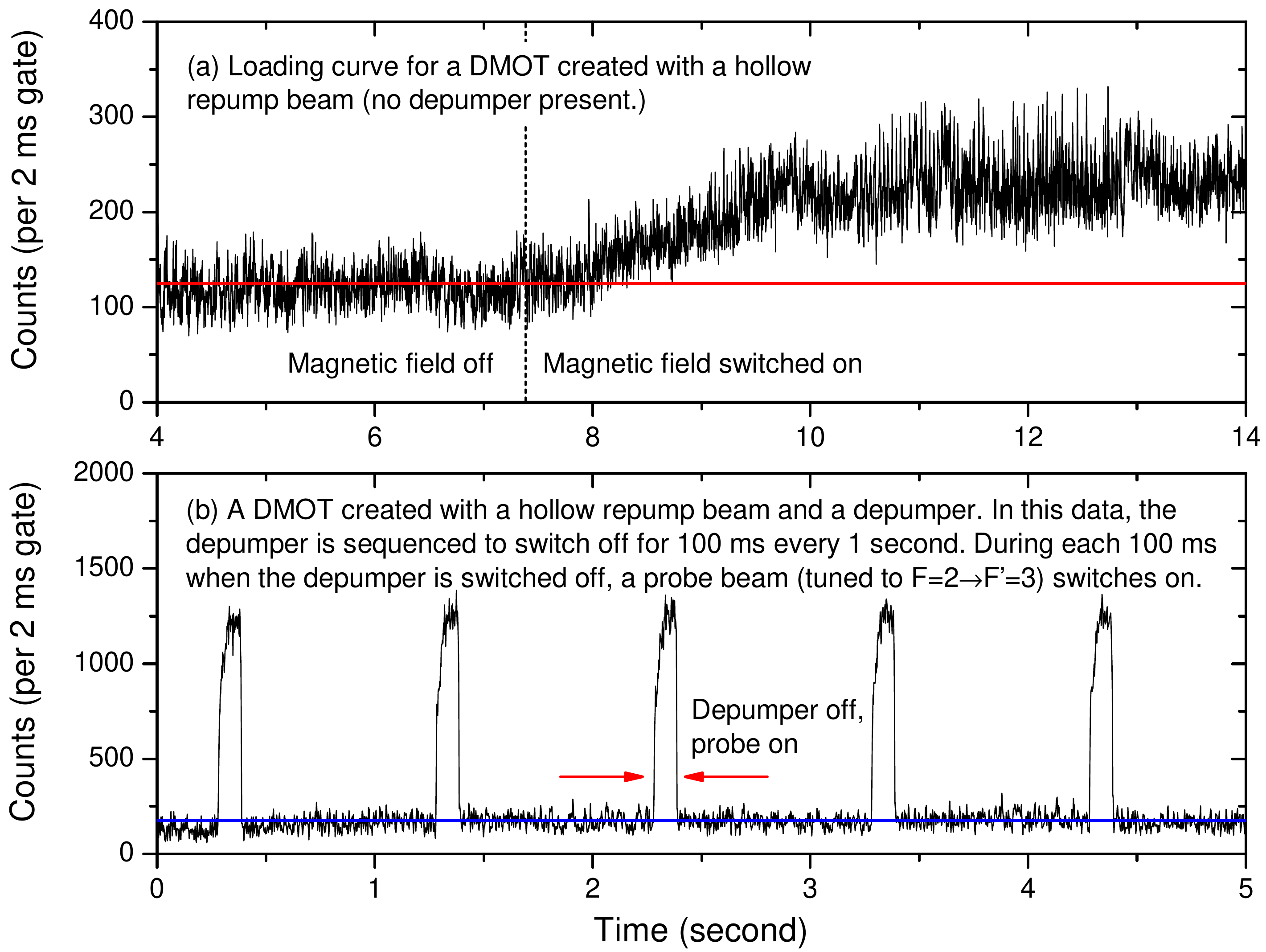}
\caption[Florescence count rates coupled into the ONF for the DMOT]{\label{fig:figure7}Florescence count rate coupled into the ONF for : (a) a DMOT created with a doughnut repump beam and no depump light, (b) a DMOT created with a doughnut repump beam and the depump beam. The depumper is switched off for 100 ms at 1 Hz repetition rate. Additionally, there is also a probe beam that switches on when the depump is off. For both plots, the doughnut-shaped repump beam is on at all times. By taking the ratio of the baseline counts in plot a to the peaks in plot b the value for $p$ is determined to be $\sim$0.10. However, by including the depumper, $p$ is determined using the baseline of plot b(blue line) to be $50/1150\approx0.04$.}
\end{figure}

\Fref{fig:figure7}(a) shows the fluorescence count rate through the fibre for a DMOT created with a doughnut repump beam (and no depumping beam). The black, vertical line at $\sim$7.5 s indicates the point at which the magnetic field is switched on to allow the DMOT to load. The fluorescence level increases from 125 counts/2 ms (indicated by the red line that shows the background light coupling into the fibre) to $\sim$275 counts/2 ms due to the residual bright atoms in the DMOT. In other words, the $F=3$ atoms contribute $\sim$150 counts/2 ms to the signal. This is the numerator of $p$ as given in \eref{eqn:equation1}. In \fref{fig:figure7}(b), the fluorescence signal is recorded for a DMOT created with the doughnut repump beam and the depumper. This yields a baseline count rate of 175 counts/2 ms (blue line, \fref{fig:figure7}(b)) and the numerator of $p$ in this case is 50 counts/2 ms. Furthermore, in this plot, the probe beam is added to the optical configuration. This probe beam is tuned to the repumping transition of $^{85}$Rb ($F=2\rightarrow F'=3$) and aligned with the dark region of the cloud. Thus, it ``fills in'' the dark central region of the doughnut-shaped repump beam, thereby recreating a standard BMOT configuration. The probe switches on for 100 ms at 1 Hz repetition rate, while the depump switches off at the same time and with the same rate. The observed spikes in the fluorescence count rate ($\sim$1150 counts/2 ms) coupled into the ONF are from atoms in both the ground state hyperfine levels, $F=2$ and 3. This is the denominator of $p$. By taking the ratio of the baseline (125 counts/2 ms) in \fref{fig:figure7}(a) to the peak counts in \fref{fig:figure7}(b), and correcting for the background level, the value for $p$ is determined as $\sim$0.10 in the DMOT without the depumper. However, with the inclusion of the depumper the dark MOT is improved and a value of $p\sim50/1150 \approx 0.04$ is obtained.

\subsection{DMOT loading times}

\begin{figure}[!ht]
\centering
\includegraphics[width=15cm]{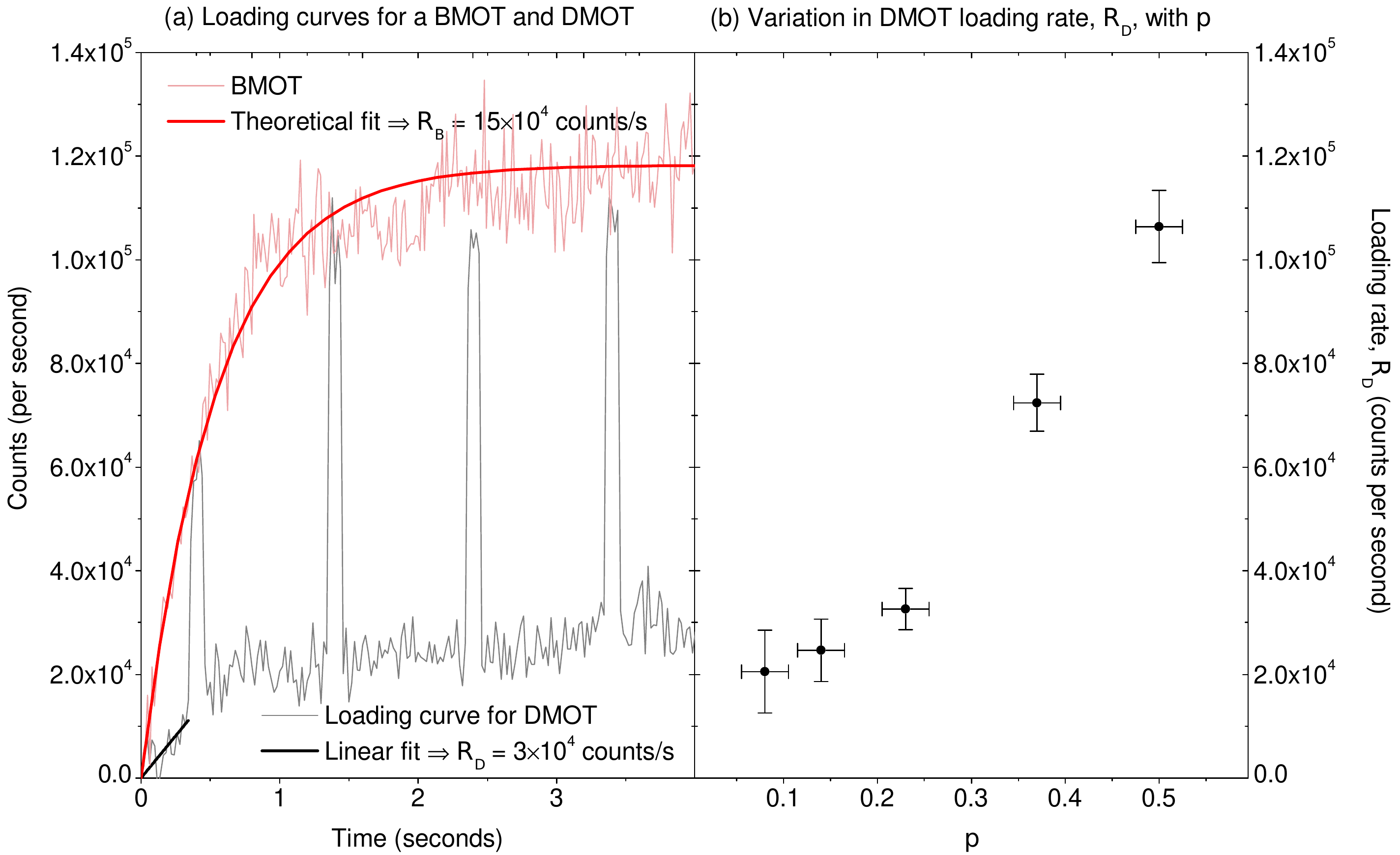}
\caption[Loading curves and loading rates for the BMOT and DMOT]{\label{fig:figure8}Background-corrected loading curves for the BMOT and DMOT are shown in (a) with theoretical fits to determine the loading rates: $R_B=15\times10^4$ counts/s, $R_D=3\times10^4$ counts/s. In this case, $p=0.22$ for the DMOT. (b) As the DMOT becomes darker (i.e. as $p$ reduces) the loading rate $R_D$ decreases.}
\end{figure}

Loading curves were analyzed for the DMOT and BMOT via the ONF with a methodology similar to that described in \cite{anderson94}. \Fref{fig:figure8}(a) shows a loading curve for a BMOT (red plot) and a loading curve for a DMOT (black plot). Both curves have been background-corrected. The BMOT curve yields a loading rate, $R_B$, of $15\times10^4$ counts/s by fitting it with $N(t)=R_B/\Gamma(1-e^{-\Gamma t})$ where $\Gamma$ is the collisional loss rate in the MOT. The plot in the lower panel represents the loading for a DMOT which has been created with a doughnut repump beam and a depumper. The spikes in fluorescence occurring at 1 Hz (for 100 ms duration) are due to  simultaneously switching off  the depumper and switching on the probe beam for detection purposes. As discussed previously, these conditions (depumper off and probe on) recreate the conditions for a BMOT and provide a measurement of the denominator of $p$. These peaks in the DMOT plot reach values of $\sim1\times10^5$ counts/s, matching the steady state count rate of the BMOT loading curve. The baseline curve of the DMOT plot represents the loading  of the fraction of $F=3$ atoms present in the DMOT. The DMOT does not contribute to the loading process so the loading rate cannot be larger in the DMOT than in the BMOT. In \fref{fig:figure8}(a) one can see that the loading rate of the DMOT, $R_D$, is $\sim0.2R_B$. For this DMOT, $p$=0.22.

\begin{figure}[!ht]
\centering
\includegraphics[width=12cm]{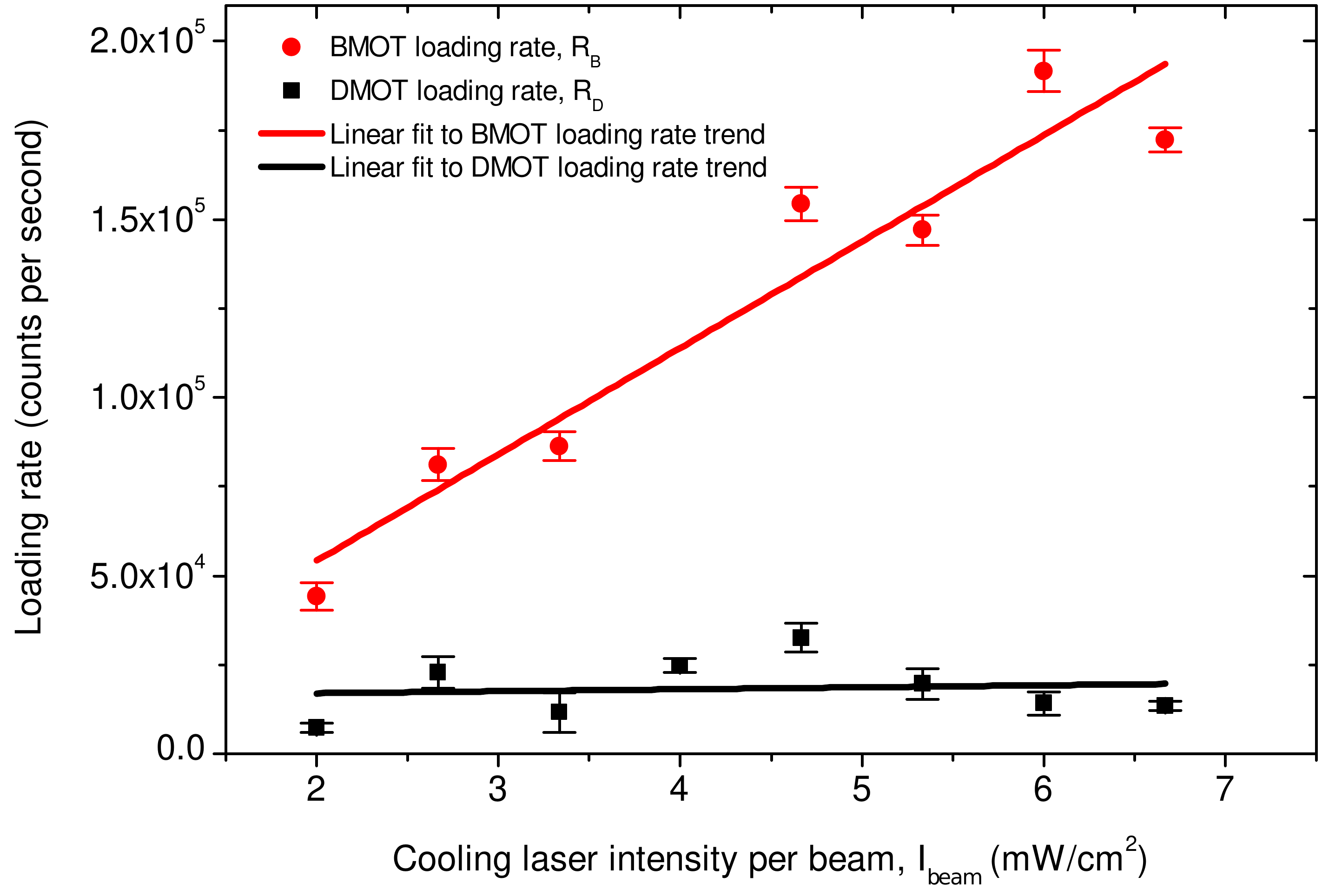}
\caption[Loading rates for a BMOT and DMOT]{\label{fig:figure9}The red (black) data points show the trend in loading rates for the BMOT (DMOT) while varying the cooling laser intensity per MOT beam, $I_{beam}$. As the doughnut repump beam and the depumper do not help the loading process, the loading rates for the DMOT are lower than those of the BMOT. Linear fits have been applied to each series as a guide to the eye. The error bars are calculated from the loading curve fits.}
\end{figure}

The change in $R_D$ can be examined as a function of $p$. In \fref{fig:figure8}(b), $p$ is varied (using different depumper intensities) over the range 0.08 to 0.50. The loading rate of the DMOT decreases for lower $p$ values as expected \cite{kim01}. Loading rates for the BMOT and DMOT can be explored by varying the cooling laser intensity (per MOT beam, $I_{beam}$) and recording the same data as shown in \fref{fig:figure8}(a) for each $I_{beam}$. This study is shown in \fref{fig:figure9}. The DMOT loading rate, $R_D$, is consistently lower than the BMOT loading rate $R_B$.

\section{Conclusion}

This work examined the implementation of a DMOT in order to circumvent the density-limitations of a BMOT. For ``atom - nanofibre'' experiments, this is a major consideration. If one wishes to do absorption experiments using an ONF, for example, signal quality will be improved dramatically by forcing as many atoms as possible to fill the evanescent region around the fibre. Although there will still be collisional losses in a DMOT (unless $p=0$), the reabsorption of scattered photons is no longer a major problem. By packing more atoms closer to the fibre surface, absorption of the light passing through the nanofiber can be enhanced as number of atoms in the evanescent field region is increased. A density improvement is critical to achieve high optical densities, thereby allowing demonstrations of nonlinear effects such as electromagnetically induced transparency and slow light \cite{braje03,hakuta08}. The variation in DMOT loading rate, $R_D$, for changing $p$ was measured via the ONF, with the lowest values of $R_D$ obtained for the darkest DMOTs as expected. This alternative technique for measurement of DMOT parameters using an ONF has the advantage that it can be used to probe different local regions of a DMOT. For example, the loss of dark state cold atoms due to collisions around the central region and near the interface between bright and dark regions of a DMOT can be studied using an ONF.  This would be extremely difficult to achieve using other commonly employed methods of fluorescence detection, i.e., using photodetectors.

\section{Ackowledgements}

This work was supported by Science Foundation Ireland under Grant No. 08/ERA/I1761 through the NanoSci- E+ Transnational Programme, NOIs, and OIST Graduate University. LR acknowledges support from IRCSET through the Embark Initiative.

\end{document}